\documentclass[12pt]{article}

\usepackage{graphics}
\usepackage{graphicx}
\usepackage{epsfig}
\usepackage{subfigure}

\usepackage{verbatim}
\usepackage{amsmath}
\usepackage{amssymb}
\usepackage{color}
\usepackage{bm}  
\usepackage{hyperref}
 \usepackage{stackrel}
 \usepackage{upgreek}
 \usepackage{csquotes}
 \usepackage{ulem}

\DeclareGraphicsExtensions{.pdf,.eps,.png,.jpg,.mps}

\newcommand{\SImum}{\ensuremath{\upmu}\textrm{m}\,}

\newcommand{\fraz}{\displaystyle\frac}
\def\tond#1{\left(#1\right)}

\def\quadr#1{\left[#1\right]}

\def\mod#1{\left|#1\right|}
\newcommand*\diff{\mathop{}\!\mathrm{d}}
\def\calE{{\cal E}}

\def\##1{{\bf #1}}
\def\=#1{\underline{\underline #1}}

\def\ux{\hat{\#x}}
\def\uy{\hat{\#y}}
\def\uz{\hat{\#z}}

\def\etao{\eta_{\scriptstyle o}}
\def\lambdao{\lambda_{\scriptstyle o}}
\def\lambdac{\lambda_{\rm c}}
\def\Nr{N_{\rm r}}
\def\Nmax{N_{\rm max}}
\def\Np{N_{\rm p}}
\def\muc{\mu_{\rm ch}}

\def\Lsub{L_{\rm sub}}

\def\epssub{\varepsilon_{\rm sub}}

\def\sigmagr{\sigma_{\rm gr}}

\def\kB{k_{\rm B}}

\begin{document}
 
\begin{center}

\textbf{Graphene pixel-based polarization-insensitive metasurface for  {almost perfect and wideband} terahertz absorption}\\

\textit{ {Pankaj Kumar,$^1$}
 {Akhlesh Lakhtakia,$^{2,3}$} and
 {Pradip K. Jain$^{1,3}$}}\\

$^1${Department of Electronics and Communication Engineering, National Institute of Technology Patna, Patna 800005, Bihar, India}\\
 $^2${NanoMM---Nanoengineered Metamaterials Group,
Department of Engineering Science and Mechanics, The Pennsylvania State University, University Park, Pennsylvania 16802, USA}\\
$^3${Material Architecture Center and Department of Electronics Engineering, Indian Institute of Technology (BHU), Varanasi 221005, Uttar Pradesh, India}\\

\end{center}

\begin{abstract}
Absorption of terahertz waves by a metasurface comprising a biperiodic array of pixellated   meta-atoms on top of a dielectric substrate backed by a perfect electric conductor was simulated using a commercial software, with either all or a few of the pixels in every meta-atom   patched with graphene. Absorptances as high as 0.99 over a 1.16-THz-wide spectral regime for normal incidence were found for a metasurface comprising meta-atoms with only a few pixels patched with graphene, regardless of the polarization state. In comparison to metasurface absorbers comprising two graphene layers separated by an insulator mounted on a metal-backed substrate, graphene need was thereby reduced by two thirds. The number and dimensions of the pixels in a meta-atom can be altered to fit spectral requirements.
 \end{abstract}

\section{Introduction}

The terahertz (THz) spectral regime---in particular, the THz gap (0.3-30 THz)---is very attractive for a variety of important applications \cite{Tonouchi}, such as imaging 
\cite{Yildrim}, spectroscopy \cite{Qin}, and cancer detection 
\cite{Yngvesson}. Accordingly, attention is being paid to the development of systems to launch, polarize, deflect, and absorb THz waves. However, conventionally designed absorbers at these frequencies have many limitations, such as large size, high operating voltage, and inflexible structure. These limitations have motivated researchers to explore new avenues for the absorption of THz waves.

Metasurfaces are being used in absorbers. For microwave frequencies, metasurface absorbers comprise periodic metallic patterns printed on top of a substrate backed by a metallic sheet deliver absorptances as high as 0.99 \cite{Ghosh}. For terahertz frequencies, metasurface absorbers comprise periodic arrays of ultrathin patches of materials such as graphene \cite{Thong}, titanium nitride \cite{Li}, gallium arsenide \cite{Ying}, and gold \cite{Liu-gold} printed on top of a metal-backed substrate.

Research on graphene-based THz metasurfaces has become popular because of simplicity, economy, and efficiency.  Graphene, a planar layer of carbon atoms bound in hexagonal structure, is the two-dimensional  version of graphite. It is a zero-bandgap material whose frequency-dependent surface conductivity  $\sigmagr$ can be dynamically controlled by varying its chemical potential $\muc$ (also called the Fermi level) either with the application of a quasistatic electric field \cite{Depine,Wang} or a magnetic bias field \cite{Wright}.  Substitutional doping can also be used to fix the chemical potential \cite{Castro}. As $\muc$ can be controlled dynamically, so can the spectra of $\sigmagr$ and the metasurface absorptance. Although graphene by itself is a modest absorber in the low-THz regime \cite{Kaipa}, graphene-containing multilayers \cite{Amin}, graphene-decorated periodically undulating surfaces \cite{Ferreira}, and graphene metasurfaces \cite{Dong} can deliver high absorptances in the THz gap.

The unit cell of a metasurface is called a meta-atom. The bottom of a meta-atom is a metal-backed dielectric substrate. Two graphene patches separated by an insulator are fabricated on the exposed face of each meta-atom in a typical graphene-based metasurface absorber \cite{Huang,Lakhtakia}. Each graphene patch in the upper layer is continuous and covers almost all of the top area of the meta-atom except for a thin frame to isolate the meta-atom from its neighbors. Each graphene patch is thus a square of side $a-d$, where $a$ is the lattice period and $d/2$ is thickness of the frame such that $d\ll{a}$.
Each graphene patch in the lower layer is a square of side $a$,  the lower layer being continuous.

\begin{figure}[htb]
\centering{\includegraphics[width=0.95 \columnwidth]{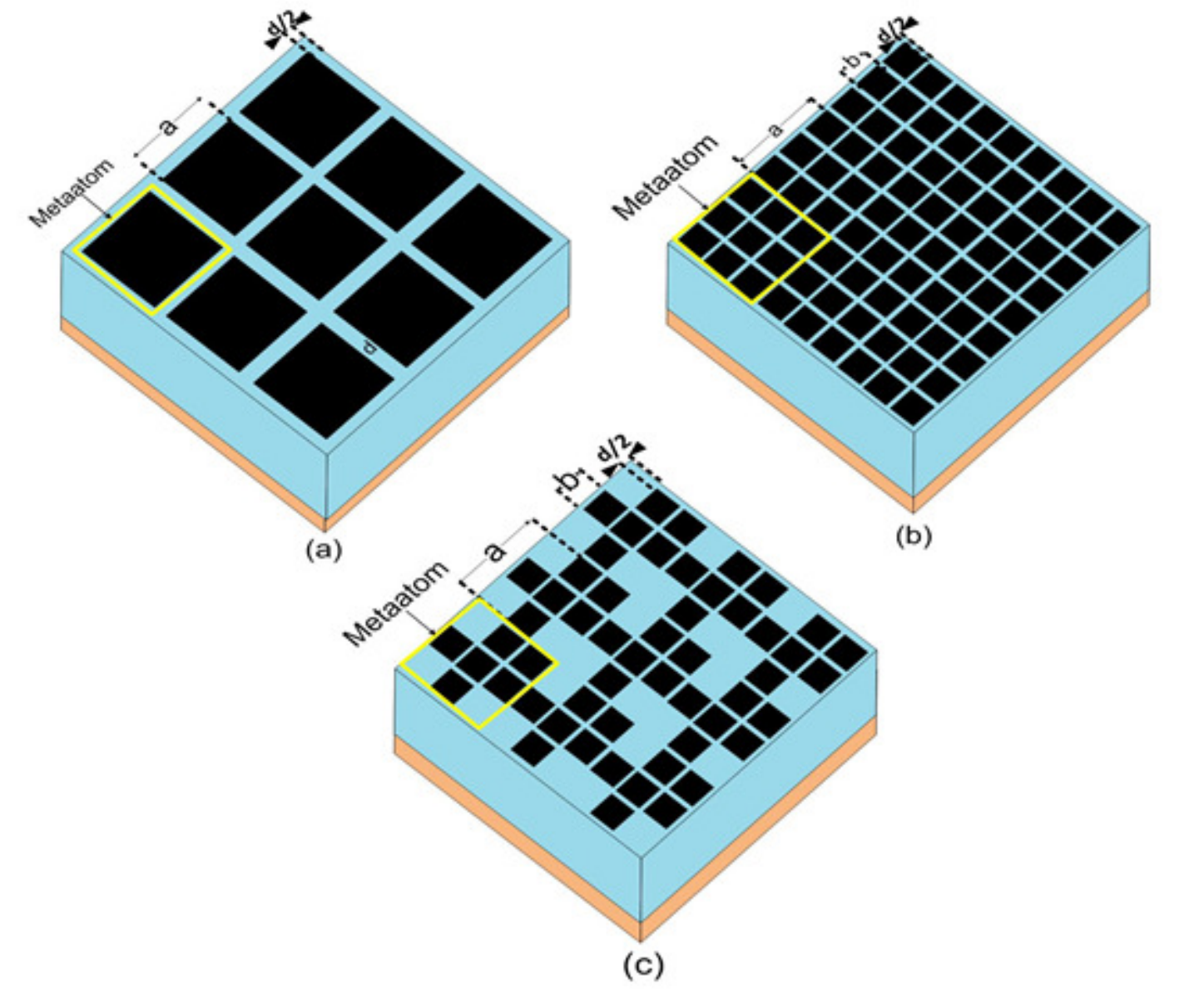}}
\caption{Top view of the meta-atoms of   absorbing metasurfaces containing graphene on top of a metal-backed substrate. (a) The meta-atom of transverse dimensions $a{\times}a$ has a graphene patch of transverse dimensions $(a-d)\times(a-d)$, $d\ll{a}$. (b) Meta-atom with graphene patch broken up into an $\Nr \times \Nr$ array of graphene-patched pixels of side $b$. (c) Meta-atom with the number $\Np$ of graphene-patched pixels less than the maximum $\Nmax=\Nr^2$.
 }
\label{Fig1}
\end{figure}

With the objective of designing a low-cost graphene-based metasurface absorber,  we decided to remove the lower graphene layer and the insulator from the typical graphene-based metasurface absorber \cite{Huang,Lakhtakia}. The resulting meta-atom is shown in Fig.~\ref{Fig1}a. As is already known \cite{Huang}, the absorption performance of the metasurface containing meta-atoms of this type is inadequate. Therefore, we adopted a  pixel-based approach \cite{Lakhtakia,Clemens,PKumarSPIE} and replaced the $(a-d)\times(a-d)$ graphene patch in Fig.~\ref{Fig1}a by $\Nmax=\Nr^2$ graphene-patched pixels, each a square of side $b$, as shown in Fig.~\ref{Fig1}b for $\Nr=3$. In order to further reduce graphene use and obtain even better absorption characteristics, we then removed some of $\Nmax$ graphene-patched pixels, leaving just $\Np <  \Nmax$ graphene-patched pixels in every meta-atom, as shown in Fig.~\ref{Fig1}c. The selection of pixels to be removed led to optimization for maximal absorptance over a wide spectral band.

\section{Simulation Details}

\begin{figure}[htb]
\centering{\includegraphics[width=0.95 \columnwidth]{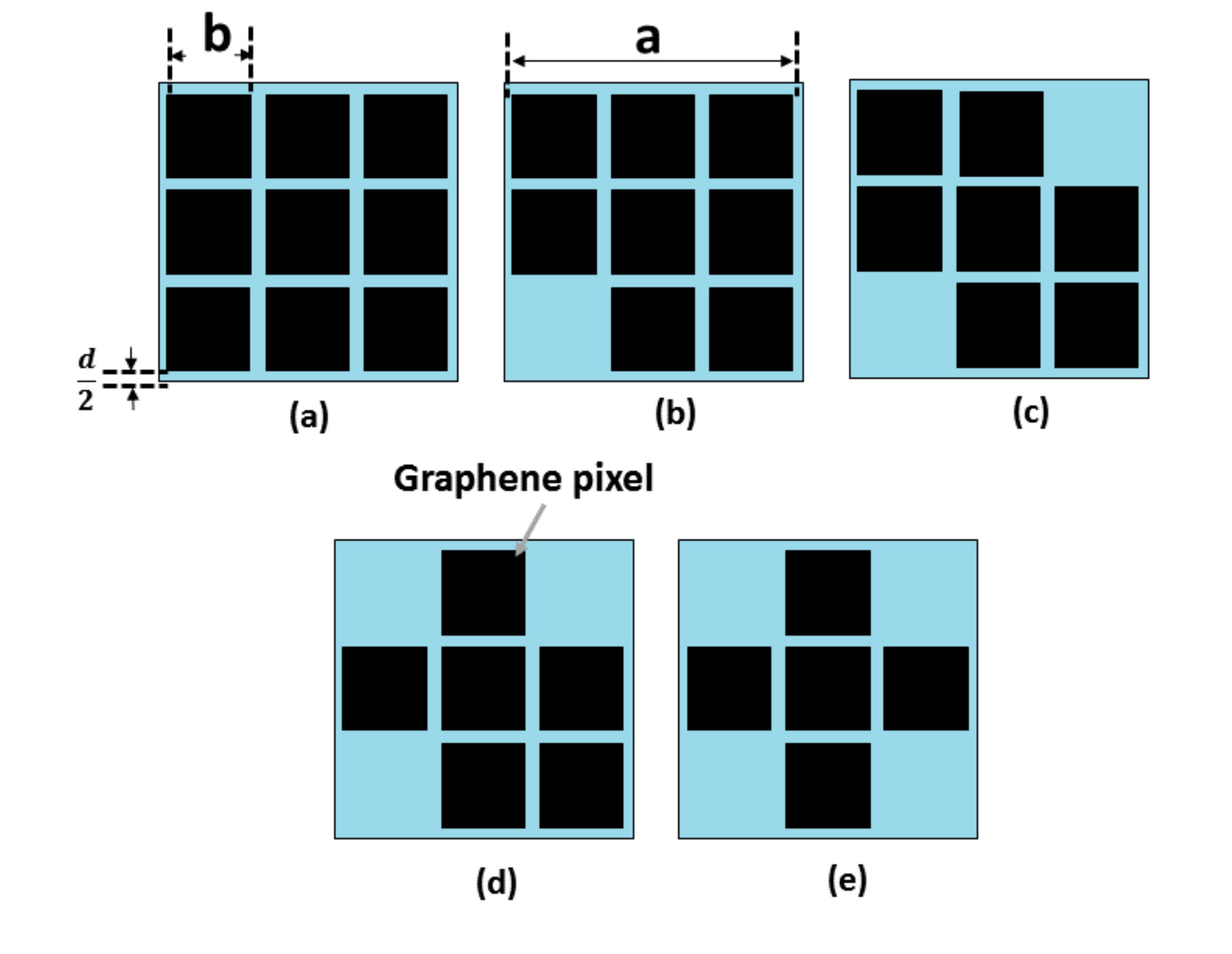}}
\caption{Top views of meta-atoms comprising $\Nmax=9$ pixels of which $\Np$ pixels are graphene-patched. (a) $\Np = 9$, (b) $\Np = 8$, (c) $\Np = 7$, (d) $\Np = 6$, and (e) $\Np = 5$.
 }
\label{Fig2}
\end{figure}

The meta-atom dimension $a<\lambdac/4$, where $\lambdac$ is the lowest value of the free-space wavelength $\lambdao$ for operation. Every pixel is separated from its nearest neighbor on every side by a thin strip of thickness $d\ll{b}$. The dimensions $b$ and  $d$ must be selected so that the ratio $\Nr=a/(b+d)$ is an integer. Of the $\Nmax$  pixels, graphene is patched on $\Np\in[1,\Nmax]$ pixels but not on the remaining $\Nmax-\Np$ pixels. Therefore, the number of unique meta-atoms realizable for a fixed number $\Np\in[1,\Nmax-1]$ of graphene-patched pixels cannot exceed $\Nmax!/\left[2\left(\Np!\right)\left(\Nmax-\Np\right)!\right]$. Figure~\ref{Fig2} presents top views of some of the meta-atoms considered by us for different values of $\Np$ when $\Nr=3$.

We took the substrate of thickness $\Lsub$ to be made a homogeneous dielectric material of relative permittivity $\epssub$. Frequency-domain simulations of the plane-wave response of the device in the 1--6-THz spectral regime (i.e., $\lambdac = 50$~$\SImum$) were carried out on the commercial 3D full-wave simulation code CST Microwave Studio\texttrademark, with the metal at the back of the substrate replaced by a perfect electrical conductor (PEC) and the temperature $T$ fixed at 300~K. In CST Microwave Studio\texttrademark,
graphene is modeled as a  0.335-nm-thick layer with
\begin{eqnarray}
\nonumber
&&\sigmagr=-\fraz{q_e^2 \tau\tond{1-i\omega\tau}}{\pi{\hbar^2}}
\Bigg[\displaystyle\int_{-\infty}^{+\infty}\fraz{\mod{\calE}}{\tond{1-i\omega\tau}^2}\fraz{\partial F\tond{\calE,\muc}}{\partial\calE}\diff\calE
\\[10pt]
&&\qquad
+\displaystyle\int_{0}^{+\infty}\fraz{F\tond{\calE,\muc}-F\tond{-\calE,\muc}}{\tond{1-i\omega\tau}^2+4\tau^2\calE^2/\hbar^2}\diff\calE\Bigg]\,,
\label{eq:sigma}
\end{eqnarray}
where $i=\sqrt{-1}$, $q_e$ is the electron charge, $\tau$ is the momentum relaxation time 
that is assumed to be independent of the energy $\calE$,   
$\hbar$ is the reduced Planck  constant, $\omega$ is the angular frequency, and
the Fermi--Dirac distribution  function
	\begin{equation}
	F\tond{\calE,\muc}=\quadr{1+\exp\tond{\fraz{\calE-\muc}{{\kB} T}}}^{-1}
	\label{eq:FermiDirac}
	\end{equation}
contains  ${\kB}$ as the Boltzmann constant. As there is provision to change $\muc$ and   $\tau$ in the software, we considered $\muc \in\left\{0,0.1,0.2,0.3\right\}$~eV and $\tau\in\left\{0.1,0.2,0.3\right\}$~ps. The dimensions $a$, $b$, $d$, and $\Lsub$ were optimized by parametric sweeps.

\indent The pixels were aligned with the $x$ and $y$ axes for the simulations, and the 2D Floquet model was implemented because of periodicity along these axes. A mesh with 21,727 tetrahedra was created.  Transmission being prevented by the PEC backing, the reflected electromagnetic field was calculated when the pixelated metasurface was irradiated by a linearly polarized plane wave propagating in the $xz$ plane at an angle  $\theta\in[0^\circ,90^\circ)$ with respect to the $z$ axis. The incident electric field phasor is given by
\begin{eqnarray}
\nonumber
\#E_{\rm inc} &=& \left[   \left(-\ux\cos\theta+\uz\sin\theta\right) \cos\varphi+\uy\sin\varphi \right]
\\
&&\qquad\times
\exp\left[i\left({2\pi}/{\lambdao}\right)\left(x \sin\theta+z\cos\theta\right)\right]\,
\end{eqnarray}
and the incident magnetic field phasor by
\begin{eqnarray}
\nonumber
\#H_{\rm inc} &=&  \etao^{-1}\left[ \left(-\ux\cos\theta+\uz\sin\theta\right) \sin\varphi- \uy\cos\varphi \right]
\\
&&\qquad\times
\exp\left[i\left({2\pi}/{\lambdao}\right)\left(x \sin\theta+z\cos\theta\right)\right]\,,
\end{eqnarray}
where the $\exp(-i2\pi{ft})$ dependence
on time $t$ is implicit with $f=\omega/2\pi$ as the  frequency,
$\etao$ is the intrinsic impedance of free space, and $\varphi\in[0^\circ,90^\circ]$ is  the polarization angle. The plane wave is transverse-magnetic (TM)  or parallel polarized when $\varphi=0^\circ$
and transverse-electric (TE) or perpendicularly polarized when $\varphi=90^\circ$.
 A post-processing module was used to compute the absorptance $A$ as a function of $\lambdao$.

\section{Numerical Results and Discussion}
\begin{figure}[htb]
\centering{\includegraphics[width=0.90 \columnwidth]{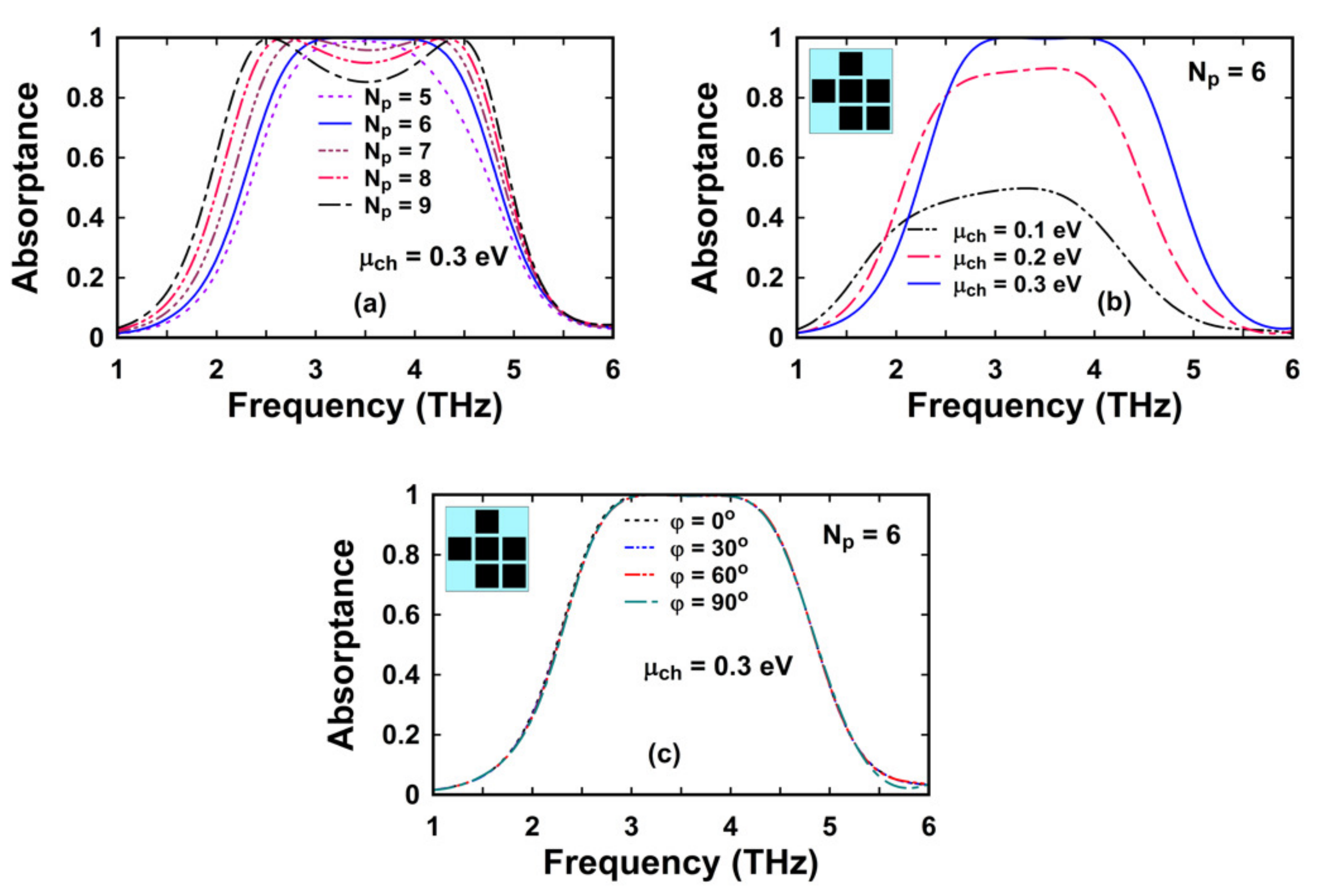}}
\caption{(a)  Absorptance spectra of the metasurfaces containing the  meta-atoms shown in Fig.~\ref{Fig2} for $\Np\in\left\{5,6,7,8,9\right\}$ when $\muc=0.3$~eV, 
$\theta=0^\circ$, and $\varphi=0^\circ$.  
(b,c) Absorption spectra for $\Np=6$ (Fig.~\ref{Fig1}d), when 
(b) $\theta=0^\circ$,  $\varphi=0^\circ$, and
  $\muc\in\left\{0.1,0.2,0.3\right\}$~eV; and
(c) $\theta=0^\circ$,  $\varphi\in\left\{0^\circ, 30^\circ,60^\circ,90^\circ\right\}$, and
  $\muc=0.3$~eV. All calculations were made with  $\tau=0.1$~ps, $T=300$~K, and $\epssub=5$.  
\label{Fig3} 
}
\end{figure}

A few examples to illustrate the pixel-based approach are presented in Figs.~\ref{Fig3} and \ref{Fig5}--\ref{Fig7} for  $\Nr=3$, $a=9.6$~\SImum, $b=3$~\SImum, $d=0.2$~\SImum, and $\Lsub=9.6$~\SImum. Whereas $\epssub=5$ is fixed in Figs.~\ref{Fig3}, \ref{Fig5}, and \ref{Fig6},
 $\epssub\in[4,5.5]$  in Fig.~\ref{Fig7}.

Absorptance spectra of the metasurfaces containing the  meta-atoms of Figs.~\ref{Fig2}a--e are
presented in Fig.~\ref{Fig3}a for $\muc=0.3$~eV, $\tau=0.1$~ps,   $T=300$~K,
$\theta=0^\circ$, and $\varphi=0^\circ$. When all pixels in a meta-atom have graphene patches 
($\Np=\Nmax$) with chemical potential $\muc=0.3$~eV, Fig.~\ref{Fig3}a shows that the absorptance
$A$ has local maxima  of $0.998$ at 2.54~THz and $0.999$ at 4.43~THz separated by a local minimum
of $0.852$ at 3.51~THz.  When the graphene patch is removed from one corner pixel (Fig. \ref{Fig2}b), the absorptance band becomes thinner but flatter with a minimum of $0.915$ at 3.51~THz. When graphene patches are removed from two pixels in diagonally opposite corners (Fig. \ref{Fig2}c), the absorptance band becomes thinner still but also flattens more, with a minimum of $0.958$ at 3.54~THz. Removal of graphene patches from three corner pixels so that $\Np=6$ (Fig. \ref{Fig2}d) makes $A\geq0.95$ for all frequencies between $2.74$ and $4.28$~THz. With graphene patches removed from all four corner pixels so that $\Np=5$ (Fig. \ref{Fig2}e), the absorptance band remains flat but has a smaller bandwidth ($A\geq0.95$) of $1$~THz. Thus, the optimal absorber contains $\Np=6$  pixels with graphene patches.

\begin{figure}[htb]
\centering{\includegraphics[width=0.95 \columnwidth]{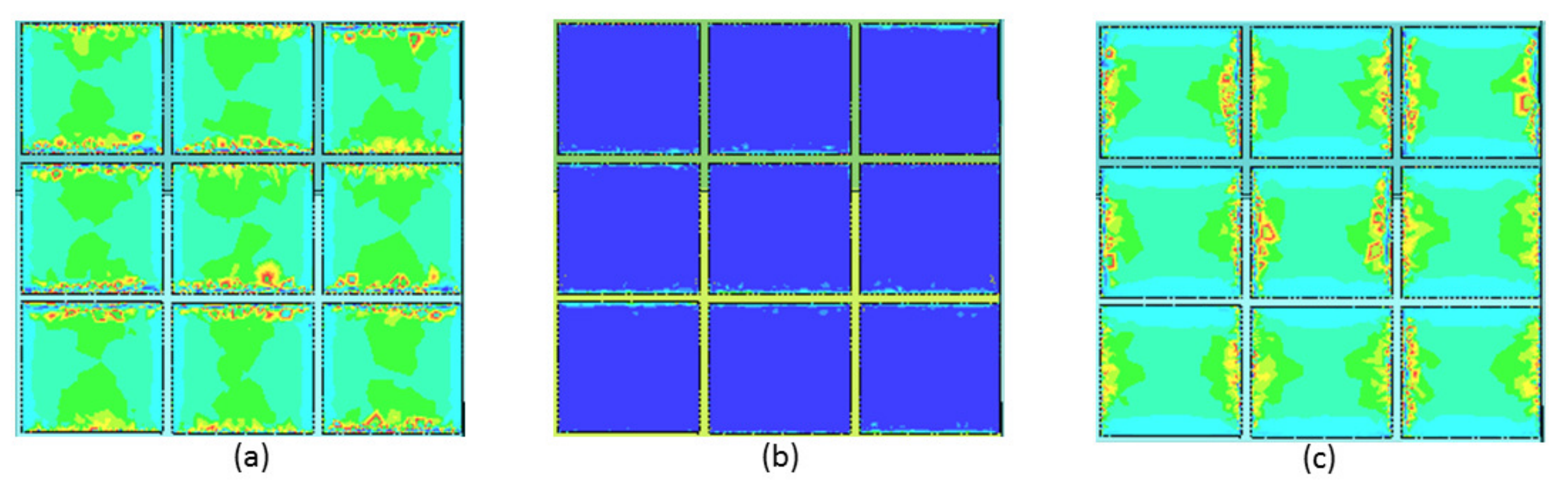}}
\caption{{Maps of the magnitude of the surface current density on the exposed surface of graphene in a meta-atom for $\Np=9$  (Fig.~\ref{Fig2}a), when
$\tau=0.1$~ps, $T=300$~K,
$\theta=0^\circ$,  $\varphi=0^\circ$, $\epssub=5$, and $\muc=0.3$~eV. (a)  $f = 2.63$~THz,  (b) $f = 3.57$~THz, and (c)  $f = 4.44$~THz.
\label{FigRes}
}}
\end{figure}

\begin{figure}[htb]
\centering{\includegraphics[width=0.95 \columnwidth]{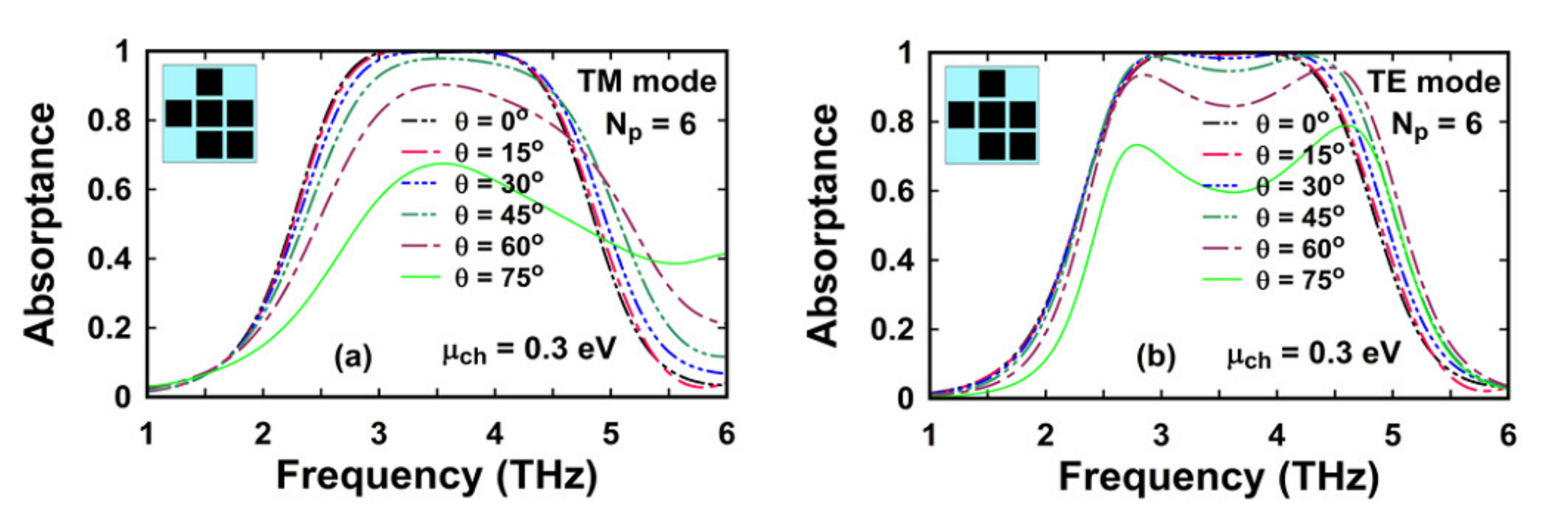}}
\caption{Absorption spectra for $\Np=6$ (Fig.~\ref{Fig2}d), when 
$\theta\in\left\{0^\circ,15^\circ,30^\circ,45^\circ,60^\circ,75^\circ\right\}$,
$\muc=0.3$~eV, $\tau=0.1$~ps,  $T=300$~K, and $\epssub=5$. (a) $\varphi=0^\circ$ and (b) $\varphi=90^\circ$.
\label{Fig5}
}
\end{figure}

\begin{figure}[htb]
\centering{\includegraphics[width=0.50 \columnwidth]{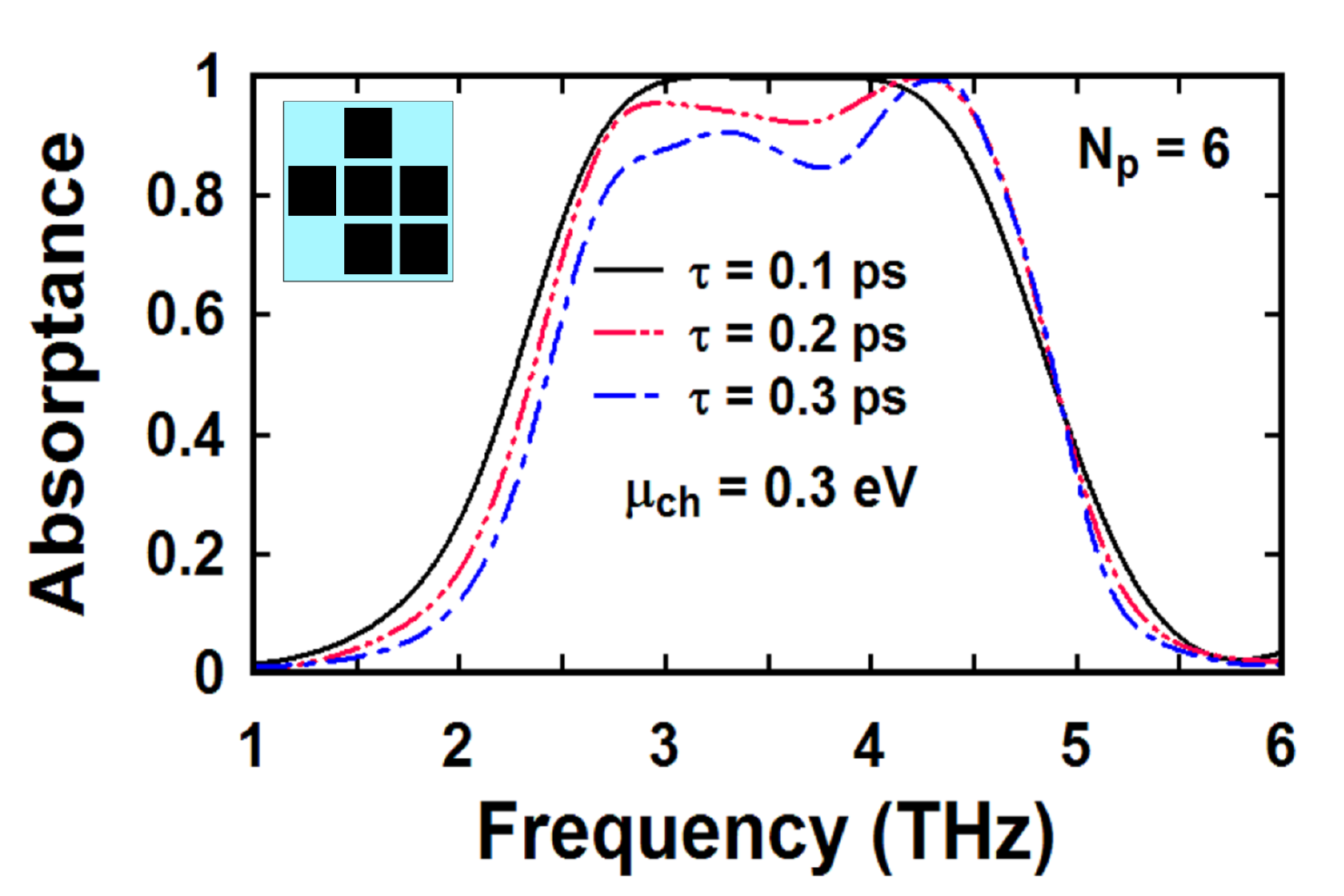}}
\caption{Absorption spectra for $\Np=6$ (Fig.~\ref{Fig2}d), when $\muc=0.3$~eV,  $\tau\in\left\{0.1,0.2,0.3\right\}$~ps, $T=300$~K,
$\theta=0^\circ$,  $\varphi=0^\circ$, and $\epssub=5$.
\label{Fig6}
}
\end{figure}

\begin{figure}[htb]
\centering{\includegraphics[width=0.50 \columnwidth]{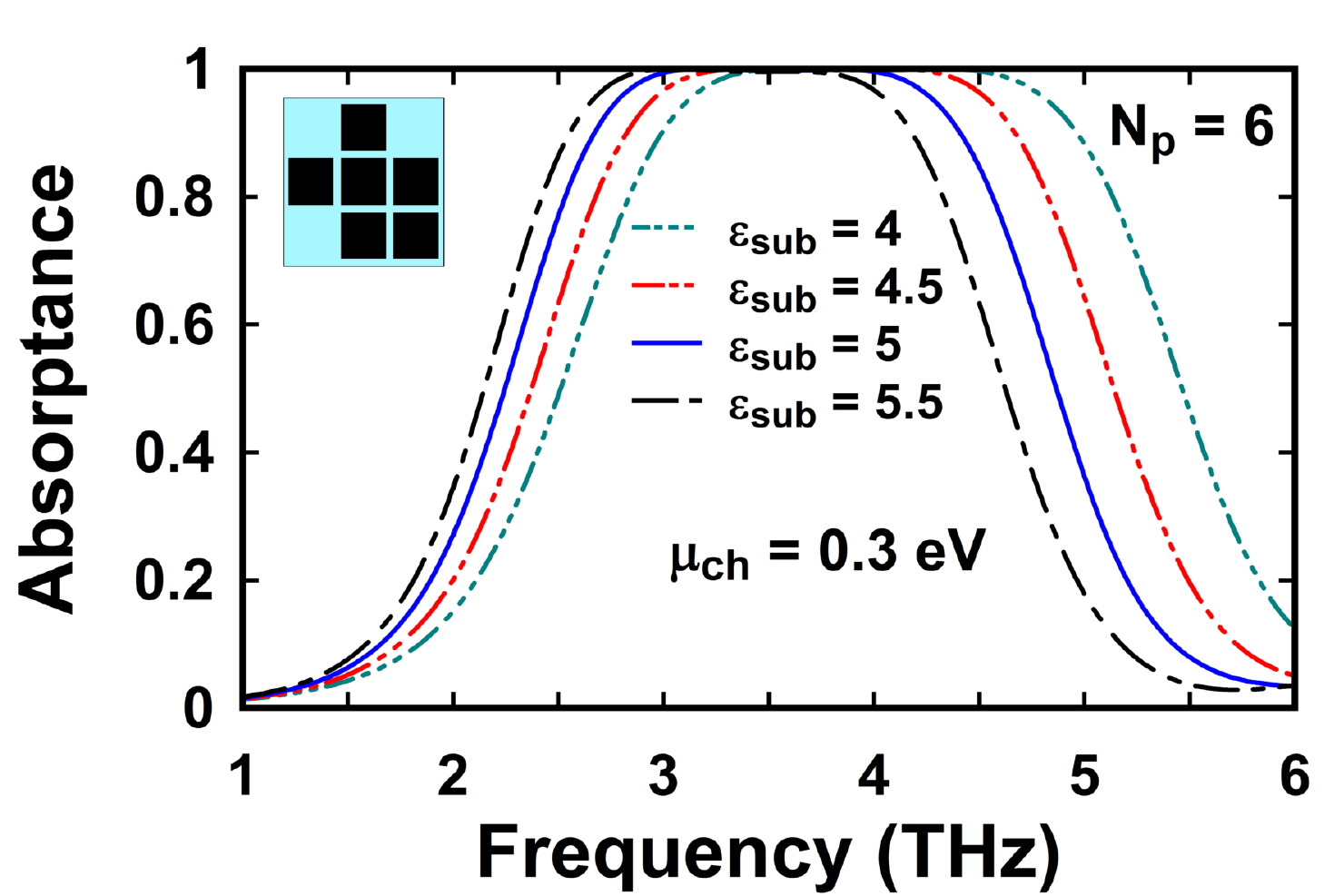}}
\caption{Absorption spectra for $\Np=6$ (Fig.~\ref{Fig2}d), when $\muc=0.3$~eV,
$\tau=0.1$~ps,   $T=300$~K,
$\theta=0^\circ$,  $\varphi=0^\circ$,
and   $\epssub\in\left\{4,4.5,5,5.5\right\}$.
\label{Fig7}
}
\end{figure}

The twin  maxima of absorptance in Fig.~\ref{Fig3}a are due to two resonances. For example, when $\Np=9$, absorptance has maxima at  $f=2.63$~THz ($A=0.991$) and
at  $f=4.44x$~THz ($A=0.996$) with an intermediate minimum   at  $f=3.57$~THz ($A=0.863$). That the two maxima are due to resonances is clear from the maps of the magnitude of the surface current density on the exposed surface of graphene at these three frequencies provided in Fig.~\ref{FigRes}. The distributions of the current density   are different for the two maxima and the magnitudes are higher
at those two maxima than at the minimum.

In an equivalent circuit model, the PEC-backed substrate acts as a capacitor in parallel with the R-L-C series circuit formed by the top array of graphene-patched pixels \cite{Huang}. As the number $\Np$ of graphene-patched pixels is reduced, the net reactance of the top array changes and the two resonances come closer to each other in Fig.~\ref{Fig3}a.

The surface conductivity of graphene in the Kubo model increases with $\muc$, and therefore it can be controlled by the application of a quasi-electrostatic field. The optimal absorber for normal incidence ($\theta=0^\circ$) and zero polarization angle (i.e., $\varphi=0^\circ$) is the metasurface whose meta-atom depicted in Fig.~\ref{Fig2}d has $\Np=6$ graphene-patched pixels. When $\muc=0.1$~eV, the absorptance spectrum in Fig.~\ref{Fig3}b has a sloping top-hat profile with $A\in[0.4,0.5]$ between
$2.04$ and $4.01$~THz. As $\muc$ increases to $0.2$~eV, the slope of the top-hat profile decreases
and $A\in[0.8,0.9]$~THz between 2.48 and 4.14~THz, as shown in  Fig.~\ref{Fig3}b. For $\muc=0.3$~eV, $A$ rises to its maximum value of unity over the  $2.92$--$4.08$-THz spectral regime.

Remarkably, the  performance of the optimal absorber is very weakly dependent on the polarization angle $\varphi\in[0^\circ,90^\circ]$ for normal incidence when $\muc=0.3$~eV. This conclusion can 
be drawn from the absorptance spectra presented in  Fig.~\ref{Fig3}c for $\theta=0^\circ$ and  $\varphi\in\left\{0^\circ, 30^\circ,60^\circ,90^\circ\right\}$.

Figure~\ref{Fig5}a presents the absorptance spectra  of the optimal absorber when the polarization angle $\varphi=0^\circ$, $\muc=0.3$~eV, $\tau=0.1$~ps, and $T=300$~K,
but the  incidence angle $\theta\in[0^\circ,75^\circ]$. As the incidence becomes more oblique, $A$ first increases and then decreases in the absorptance band. Figure~\ref{Fig5}b provides the analogous graphs for $\varphi=90^\circ$,  with $A$ decreasing
monotonically as $\theta$ increases. However, even for $\theta=45^\circ$, $A\geq0.95$ over the  $2.7$--$4.5$-THz spectral regime, whether the incident plane wave is TM polarized (Fig.~\ref{Fig5}a) or TE polarized (Fig.~\ref{Fig5}b). Thus, the optimal absorber delivers controllable, wideband, polarization-insensitive, near-unity absorptance.

The momentum relaxation time
$\tau$ can be expected to affect the results, although $\tau$ cannot be controlled after fabrication.
 The absorptance spectra  of the optimal absorber (Fig.~\ref{Fig2}d) are shown in Fig.~\ref{Fig6}  for $\muc=0.3$~eV,  $T=300$~K,  $\tau\in\left\{0.1,0.2,0.3\right\}$~ps,
$\theta=0^\circ$,  $\varphi=0^\circ$, and $\epssub=5$. Clearly, $A$
decreases as $\tau$ increases, thereby reducing the electron scattering rate and electron mobility in   graphene.

The substrate also affects device performance. Figure~\ref{Fig7} presents the absorptance spectra  of the optimal absorber  (Fig.~\ref{Fig2}d) when $\theta=0^\circ$,  $\varphi=0^\circ$,    $\muc=0.3$~eV, $\tau=0.1$~ps,   $T=300$~K, and $\epssub\in\left\{4,4.5,5,5.5\right\}$. As $\epssub$ decreases, the capacitance of the PEC-backed substrate decreases so that the resonance frequencies blueshift. Accordingly, the absorption band blueshifts as $\epssub$ decreases towards unity.

\begin{figure}[htb]
\centering{\includegraphics[width=0.95 \columnwidth]{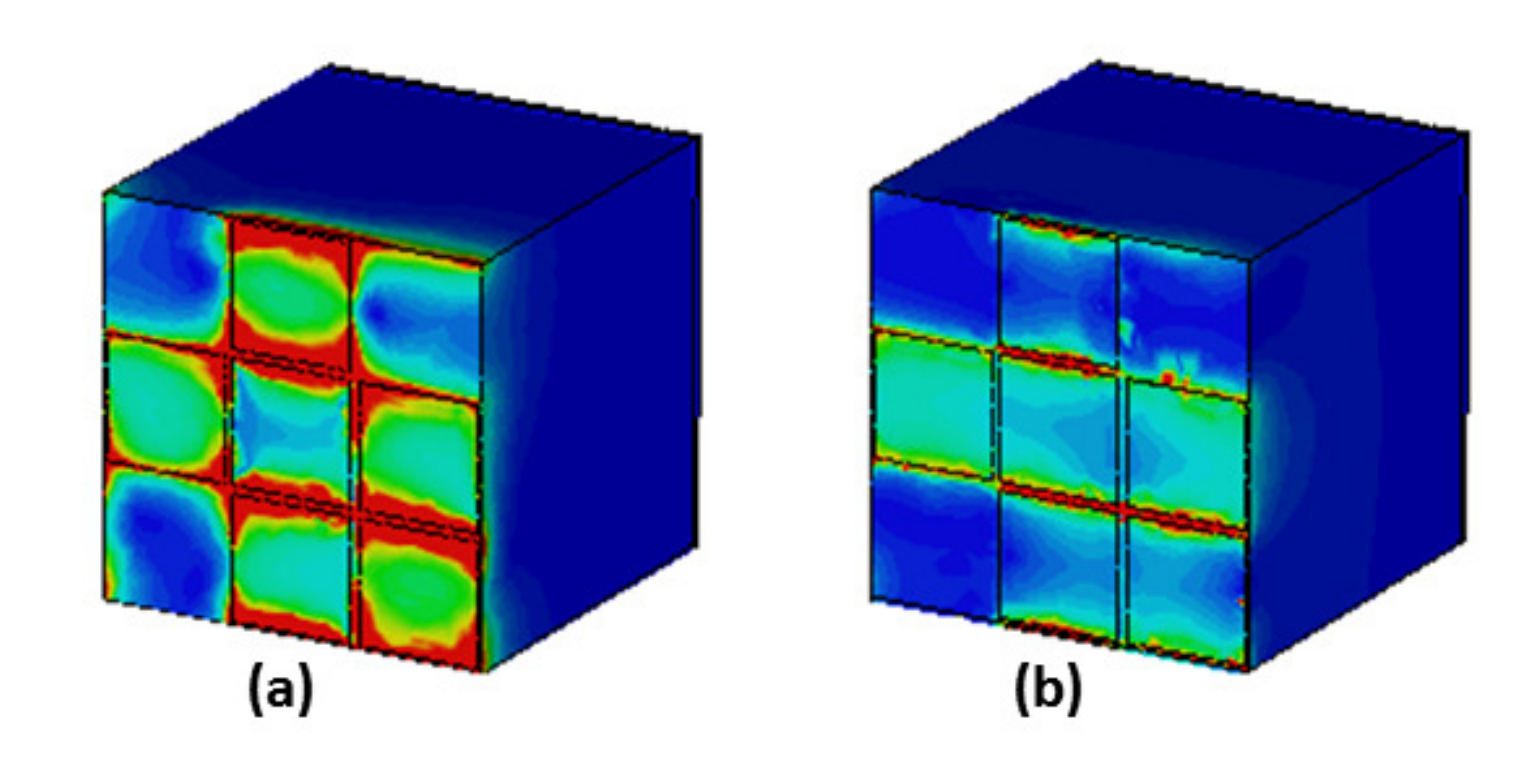}}
\caption{Maps of the magnitude of the electric field  on the exposed surface of graphene in one meta-atoms for $\Np=6$ (Fig.~\ref{Fig2}d) when
$\tau=0.1$~ps, $T=300$~K.
$\theta=0^\circ$,  $\varphi=0^\circ$, $\epssub=5$, and $f = 3.5$~THz. (a) $\muc=0.3$~eV   
 $\Rightarrow A=0.99$ and (b) $\muc=0.1$~eV  
 $\Rightarrow  A=0.49$.
\label{Fig8}
}
\end{figure}

In an attempt to explain the role of $\muc$ in enhancing $A$, we investigated the
spatial variation of the magnitude of the electric field ${\bf E}$ on the exposed surface of graphene
in the optimal absorber (Fig.~\ref{Fig2}d) at $f=3.5$~THz, when
$\tau=0.1$~ps, $T=300$~K.
$\theta=0^\circ$,  $\varphi=0^\circ$, and $\epssub=5$. According to Fig.~\ref{Fig5}b,
$A=0.99$ when $\muc=0.3$~eV, but $A=0.49$ when $\muc=0.1$~eV. Maps of
the magnitude of the electric field  on the exposed surface of graphene    
are presented in Fig.~\ref{Fig8} for $\muc\in\left\{0.1,0.3\right\}$~eV. The electric field
is much more intense at the higher value
of $\muc$ than at the lower value of $\muc$. Now, $\sigmagr= 0.00105 + 0.000368i$~S for $\muc=0.1$~eV
but $\sigmagr=0.00316 + 0.001105i$~S for $\muc=0.3$~eV at $f=3.5$~THz.  At the higher value of $\muc$, the magnitude of the electric field  is higher
in graphene and ${\rm Re}\left(\sigmagr\right)$ is also higher; hence,
 the absorption rate per unit area, $(1/2){\rm Re}\left(\sigmagr\right)   \vert{\bf E}\vert^2$, in graphene is greatly enhanced,
 leading to a very high value of $A$.

Finally, a  comparison of our results with known results for other metasurface absorbers is in order.
Batrakov \textit{et al.} \cite{Batr} fabricated a cascade of five graphene layers to obtain $A = 0.65$ at $\sim1$~THz with $\sim0.2$-THz full-width-at-half-maximum (FWHM) bandwidth. In contrast, our optimal device can have, for example, $A\geq0.95$ for $3.5\pm 0.75$ THz (Fig.~\ref{Fig2}c). Thus, our device has higher absorptance over a much wider bandwidth, and using much less graphene. Nikitin  \textit{et al.} \cite{Niki} simulated absorption in a graphene sheet with   circular perforations of 2.5-\SImum diameter arrayed on a square grid of side 5~\SImum. Resonant absorption did not exceed 0.5 with a $0.03$-THz FWHM bandwidth in the 4--15 THz spectral regime. Similar comments hold about the experimental results of Liu \textit{et al.} \cite{Liu}. The absorption bands obtained from simulations by Hajian \textit{et al.} \cite{Haj} are much narrower and the absorption peaks are lower than those delivered by the simulations in our pixel-based approach.

\section{Concluding Remarks}
Graphene-based metasurface absorbers of terahertz radiation typically employ meta-atoms, each of which is an assembly of two graphene layers, separated by an insulator, mounted on top of a 
metal-backed substrate. We eliminated one graphene layer as well as the insulator. Furthermore, we replaced the remaining graphene layer by an array of graphene patches. 
Using this pixel-based approach, we determined the geometric parameters of an optimal absorber that delivers controllable and polarization-insensitive absorptance exceeding $0.99$ over a 1.16-THz--wide spectral regime for normal incidence. This optimal absorber reduces graphene need by two thirds in comparison to approaches employing two graphene layers separated by an insulator  \cite{Huang}   and still enhances absorption to virtually the maximum possible---that too, in a wider spectral regime. The spectral regime can be either blueshifted or redshifted somewhat by suitably scaling some or all dimensions of the device \cite{Sinclair}.

\vspace{0.5cm}
\noindent {\bf Acknowledgments.}  The authors gratefully acknowledge helpful comments from an anonymous reviewer. AL and PKJ thank the Visiting Advanced Joint Research (VAJRA) program of the   Department of Science and Technology, Government of India for funding their collaborative research. AL thanks the Charles Godfrey Binder Endowment at the Pennsylvania State University for ongoing support of his research activities.

\end{document}